\documentclass[twocolumn,trackchanges]{aastex701}

\usepackage{amsmath}
\usepackage{soul}

\begin{document}

\title{ELFO: A Python Package for Emission Line Fitting Optimization in Integral Field Spectroscopy Data} 

\author[orcid=0009-0001-6324-944X]{Hui Guo} 
\affiliation{Department of Astronomy, University of Science and Technology of China, Hefei, Anhui 230026, People's Republic of China}
\affiliation{School of Astronomy and Space Science, University of Science and Technology of China, Hefei 230026, People's Republic of China}
\email{guohui@mail.ustc.edu.cn}

\author[0000-0003-4286-5187]{Guilin Liu}
\affiliation{Department of Astronomy, University of Science and Technology of China, Hefei, Anhui 230026, People's Republic of China}
\affiliation{School of Astronomy and Space Science, University of Science and Technology of China, Hefei 230026, People's Republic of China}
\email{glliu@ustc.edu.cn}

\author[0009-0006-3887-8988]{Jianghui Xu}
\affiliation{Department of Astronomy, University of Science and Technology of China, Hefei, Anhui 230026, People's Republic of China}
\affiliation{School of Astronomy and Space Science, University of Science and Technology of China, Hefei 230026, People's Republic of China}
\email{jianghuixu@mail.ustc.edu.cn}

\author[]{Chao Geng} 
\affiliation{Department of Astronomy, University of Science and Technology of China, Hefei, Anhui 230026, People's Republic of China}
\affiliation{School of Astronomy and Space Science, University of Science and Technology of China, Hefei 230026, People's Republic of China}
\email{gengchao@ustc.edu.cn}

\author[0000-0003-3667-1060]{Zhicheng He}
\affiliation{Department of Astronomy, University of Science and Technology of China, Hefei, Anhui 230026, People's Republic of China}
\affiliation{School of Astronomy and Space Science, University of Science and Technology of China, Hefei 230026, People's Republic of China}
\email{zcho@ustc.edu.cn}

\author[0000-0002-3073-5871]{Shiyin Shen}
\affiliation{Shanghai Astronomical Observatory, Chinese Academy of Sciences, 80 Nandan Rd., Shanghai, 200030, China}
\affiliation{Key Lab for Astrophysics, Shanghai, 200034, China}
\email{ssy@shao.ac.cn}

\author[]{Lei Hao}
\affiliation{Shanghai Astronomical Observatory, Chinese Academy of Sciences, 80 Nandan Rd., Shanghai, 200030, China}
\email{haol@shao.ac.cn}

\correspondingauthor{Guilin Liu}
\email{glliu@ustc.edu.cn}
\correspondingauthor{Zhicheng He}
\email{zcho@ustc.edu.cn}

\begin{abstract}

Integral field spectroscopy (IFS) provides spatially resolved spectra, enabling detailed studies that address the physical and kinematic properties of the interstellar medium. A critical step in analyzing IFS data is the decomposition of emission lines, where different velocity components are often modeled with Gaussian profiles. However, conventional fitting methods that treat each spectrum independently often yield spatial discontinuities in the fitting results. Here, we present Emission Line Fitting Optimization (ELFO), a Python package for IFS spectral fitting. ELFO uses the results of neighboring spectra to determine multiple initial guesses and selects the result that exhibits spatial smoothness. We tested ELFO on IFS data of two quasars obtained from the Multi-Unit Spectroscopic Explorer, where it successfully corrected anomalous fits, revealed previously unresolved substructures, and made large-scale kinematic structures more evident. With minor modifications, this method can also be easily adapted to other IFS data and different emission lines.

\end{abstract}
\section{Introduction}
\label{sec1}
It is widely accepted that almost all massive galaxies host a supermassive black hole (BH) at their center \citep{1998AJ....115.2285M}. Active galactic nuclei (AGN) feedback, which occurs through the interaction between the energy and radiation released during the accretion process of a supermassive black hole and the gas in the host galaxy, can significantly influence galaxy formation and evolution \citep{2012ARA&A..50..455F}. However, the large disparity in scale between galaxy and black hole makes it challenging to study the detailed processes of such feedback. A powerful tool to tackle this challenge is integral field spectroscopy (IFS). First proposed in the 1980s, IFS enables the acquisition of a spectrum for each 2D spatial pixel (spaxel) of the object in a single exposure \citep{bacon2017optical}. Therefore, IFS is particularly suitable for observing spatially extended astronomical objects and exploring AGN feedback. \citet{2015ApJS..221....9L} presented integral field unit (IFU) observations of the ionized gas around two quasars obtained from the Gemini-South telescope, and found a quasi-spherical outflow with a radius of $\sim$1.2 kpc after spectroscopically removing the tidal tail. \citet{2023SciA....9G8287S} discovered spectacular quasar-driven superbubbles in three red quasars with IFU data from Gemini-North Multi-Object Spectrographs (GMOS-N). \citet{2025ApJS..277...57Z} observed quasar 3C 191 with the Spectrograph for Integral Field Observations in the Near Infrared (SINFONI) on the Very Large Telescope (VLT)and identified a galactic-scale emission-line outflow. In addition to these instruments, the Multi-Unit Spectroscopic Explorer (MUSE) on the VLT \citep{2010SPIE.7735E..08B} and the forthcoming Chinese Space Station Telescope Integral Field Spectrograph (CSST-IFS) \citep{2011SSPMA..41.1441Z} can also provide high-resolution IFS data.

The broad-line region (BLR), narrow-line region (NLR), and ionized outflow in an AGN, although differing in scale, can all produce bright emission lines \citep{2015ARA&A..53..365N}. In this context, the decomposition of the emission line components is crucial to uncovering the physical information embedded in the spectral data. For the analysis of emission lines, it is often assumed that each velocity component follows a Gaussian profile \citep{2013MNRAS.436.2576L,2021MNRAS.500.2871S}. This assumption arises from the fact that Doppler shifts, which are caused by the random thermal motion of atoms or molecules in the gas, along with other line-broadening effects, cause the emission line profiles of single gas components to approximate a Gaussian shape.

In previous studies, several spectral fitting toolkits for IFS have been proposed \citep{2016AJ....152...83L,2016Ap&SS.361..280H,2024A&A...688A..69F}. Normally, the same parameters are applied to the spectral fitting of all spaxels \citep{2023ascl.soft10004R}. While this approach is straightforward and intuitive, it is not optimal when the data signal-to-noise ratio (SNR) is insufficient. A key issue is that the fitting results of neighboring spectra are correlated. Fitting each spectrum independently can lead to non-physical discrepancies among the results of neighboring spectra. Several techniques have been proposed to tackle the fitting challenges caused by relatively low SNR in spectra. The most widely used method is spatially binning the 3D spectral data to achieve the desired SNR \citep{2020A&A...644A..15M,2019A&A...628A.117B}. The main problem with this SNR enhancement technique is that it may artificially combine regions that do not share the same physical state. \citet{2017MNRAS.466.3989C} considered the continuity of physical processes during binning, but spatial information was still lost. \citet{2019MNRAS.482.3880G} employed spatial correlations to model the physical properties of galaxies, but did not apply these correlations to improve the fitting of the emission lines. Some previous studies attempted to utilize the spatial correlations in IFS data to improve the fitting of spectral emission lines: \citet{2016Ap&SS.361..280H} used the median of the IFS spectral fitting results as the initial guesses for further fitting. \citet{2019A&A...628A..78R} employed machine learning to determine the initial guesses and the number of Gaussian components and performed spatially coherent refitting. However, machine learning incurs a training cost. In addition, this work only takes into account neighboring spectra.

To optimize the emission line fitting results, we have developed an IFS spectral fitting algorithm \citep{guo_2025_17747778}: Emission Line Fitting Optimization\footnote{\url{https://github.com/GuloHui/csst-ifs-elfo}} (ELFO; written in Python, see more details of the code in online documentation\footnote{\url{https://csst-ifs-elfo.readthedocs.io}}), which is based on \texttt{PyQSOFit} \citep{2018ascl.soft09008G}. Our starting point is that parameters like the velocity of the gas should vary smoothly along spatial dimensions, due to the continuity of physical processes in galaxies and the limitations of spatial resolution. The initial guess for the emission line fitting of each spectrum is set based on the results of neighboring spectra, and different fitting orders yield different results. The fitting result, which is spatially smooth, is then selected from all the solutions obtained.

In this paper, we provide an overview of the principle and usage of the ELFO algorithm and show two examples of its application. In Section \ref{sec2}, we introduce the data and the spectral fitting method used in this work. Section \ref{elfo} presents a detailed description of the ELFO algorithm. The results of the two application cases, along with their interpretation, are given in Section \ref{sec4}. Finally, we summarize our main conclusions in Section \ref{sec5}.

\begin{figure*}[t]
\centering
\includegraphics[width=1\textwidth]{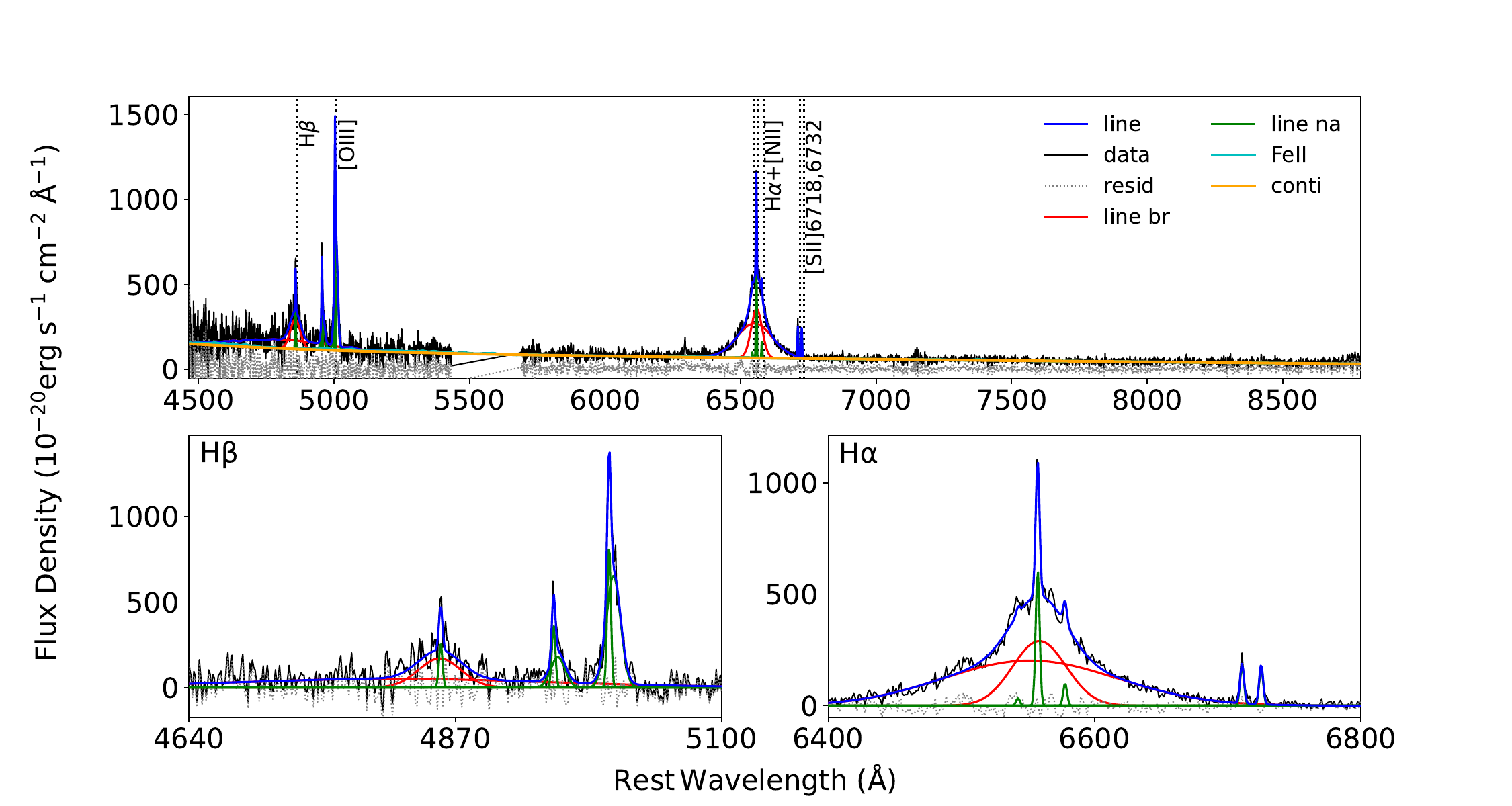}
\caption{Example of spectral fitting results. In the upper panel, the black solid line illustrates the observed data, and the blue solid line depicts the best-fitting model obtained from \texttt{PyQSOFit}. The model includes the continuum (orange solid line), Fe II emission (cyan solid line), and emission lines (red solid line for broad emission lines, green solid line for narrow emission lines). The gray dashed line represents the residuals of the spectral fitting. The bottom two panels show the emission line fitting results of the H$\beta$ and H$\alpha$ regions, respectively.}
\label{fig1}

\end{figure*}

\section{Data and Spectral Fitting}

\label{sec2}

\subsection{Data}
To test our algorithm, we employed reduced MUSE data of two nearby quasars ($z = 0.064,\ 0.060$), MR 2251-178 and PG 1126-041, obtained from the link provided by \citet[hereafter Marasco2020]{2020A&A...644A..15M}. MR 2251-178 is a radio-quiet, type 1 luminous quasar surrounded by a giant emission-line nebula. PG 1126-041, also known as Mrk 1298, is a radio-quiet AGN, with a luminosity between that of typical Seyfert galaxies and quasars. Table \ref{tab1} summarizes the main physical properties of these two quasars \citep{2020A&A...644A..15M}.

The two quasars were observed using MUSE with adaptive optics (AO) \citep{2012SPIE.8447E..37S,2008SPIE.7015E..24A} in the Narrow Field Mode (NFM). This mode covers a wavelength range of 4750-9350 Å, with a spectral resolution $R$ of about 1750 in the blue and 3600 in the red. Data for MR 2251-178 (PG 1126-041) are structured as a cube with $ 381 \times 354\ (362 \times 357)$ spaxels, covering a field of view of $9.65^{\prime\prime} \times 8.97^{\prime\prime}\ (9.17^{\prime\prime} \times 9.05^{\prime\prime})$. AO can alter the shape of the point spread function (PSF). Therefore, we adopt the full width at half power (FWHP) of PSF reported by Marasco2020 to represent the spatial resolution. The values are approximately $0.54^{\prime\prime}$ for MR 2251-178, and $0.36^{\prime\prime}$ for PG 1126-041. The difference in spatial resolution between the two datasets primarily arises from variations in seeing conditions during the observations.

\begin{table}[h]
\centering
\caption{Physical properties of the two quasars}
\fontsize{9}{15}\selectfont
\begin{tabular}{ccc}
        \hline
        & \textbf{MR 2251-178} & \textbf{PG 1126-041} \\
        \hline
        RA (J2000) & $22^\text{h}54^\text{m}05.936^\text{s}$ & $11^\text{h}29^\text{m}16.704^\text{s}$ \\
        Dec (J2000) & $-17^\circ34'53.87''$ & $-04^\circ24'06.66''$ \\
        $z$ & 0.064 & 0.060 \\
        Scale (kpc/arcsec) & 1.23 & 1.16 \\
        $\log(L_{\text{BOL}}/L_\odot)$ & $12.15 \pm 0.11$ & $11.62 \pm 0.15$ \\
        $\log(M_{\text{BH}}/M_\odot)$ & $8.43 \pm 0.43$ & $7.75 \pm 0.43$ \\
        
        \hline
    \end{tabular}

\label{tab1}
\end{table}

\subsection{Spectral Fitting}
In this study, we fit each spectrum with \texttt{PyQSOFit} \citep{2018ascl.soft09008G,2019ApJS..241...34S}. \texttt{PyQSOFit} is a software designed to decompose quasar spectra using the Levenberg-Marquardt least squares method. Initially, the spectrum is shifted to the rest frame, followed by a correction for Galactic reddening. The continuum fit is then performed in spectral windows without significant emission lines. The continuum model incorporates a power law function to represent the AGN continuum and a low-order polynomial to approximate the stellar contribution and intrinsic dust extinction. In addition to the continuum model, we use empirical templates \citep{1992ApJS...80..109B,2001ApJS..134....1V} to fit the Fe II emission simultaneously.
Subsequently, the continuum and Fe II emission models are subtracted, leaving a spectrum that contains only the emission lines.

\begin{table}[h]
\centering
\caption{Line Fitting Parameters}
\label{tab2}
\setlength{\tabcolsep}{3pt}
\fontsize{8}{11}\selectfont
\begin{tabular}{c c cc}
\hline
\textbf{Line Complex} & \textbf{Fitting Range (\AA)} & \textbf{Line} & \textbf{\(n_{\text{Gauss}}\) } \\
\hline
H$\alpha$ & 6400--6800 & Broad H$\alpha$ & 2 \\
          &            & Narrow H$\alpha$ & 1 \\
          &            & [\ion{N}{2}] 6549 & 1 \\
          &            & [\ion{N}{2}] 6585 & 1 \\
          &            & [\ion{S}{2}] 6718 & 1 \\
          &            & [\ion{S}{2}] 6732 & 1 \\
\hline
H$\beta$  & 4640--5100 & Broad H$\beta$ & 2 \\
          &            & Narrow H$\beta$ & 1 \\
          &            & [\ion{O}{3}] 4959 core & 1 \\
          &            & [\ion{O}{3}] 5007 core & 1 \\
          &            & [\ion{O}{3}] 4959 wing & 1 \\
          &            & [\ion{O}{3}] 5007 wing & 1 \\
\hline

\end{tabular}


\end{table}

The fitting is then conducted on individual spectral line complexes, each consisting of multiple lines close in wavelength. For each complex, a set of Gaussian functions is applied to fit individual lines. Table \ref{tab2} presents extensive information of line complexes and their fitting parameters \citep{2019ApJS..241...34S}. The minimum value of the narrow line width is determined according to the spectral resolution, while the maximum value is fixed at 1200 km s$^{-1}$. The narrow components of H$\alpha$, [\ion{N}{2}]\(\lambda\lambda 6549, 6585\), and [\ion{S}{2}] \(\lambda\lambda 6718, 6732\) are each modeled with a single Gaussian profile, with their velocity offsets from the systemic redshift and line widths tied to be identical. The same constraints are imposed on the narrow components of H$\beta$ and the core [\ion{O}{3}]\(\lambda\lambda 4959, 5007\) components. The flux ratio of the [\ion{S}{2}] \(\lambda\lambda 6718, 6732\) doublet is set to 1 since the lines are weak and their ratio has negligible influence on our results. The flux ratio of the [\ion{N}{2}]\(\lambda\lambda 6549, 6585\) doublet is constrained to the theoretical flux ratio \({f_{6585}}/{f_{6549}} = 3\). Figure \ref{fig1} shows an example of a spectral fitting result using \texttt{PyQSOFit}.

\begin{figure*}[t]

\centering
    \begin{minipage}{0.45\textwidth}  
        \centering
        \includegraphics[width=\textwidth,trim={0 0 0 90}, clip]{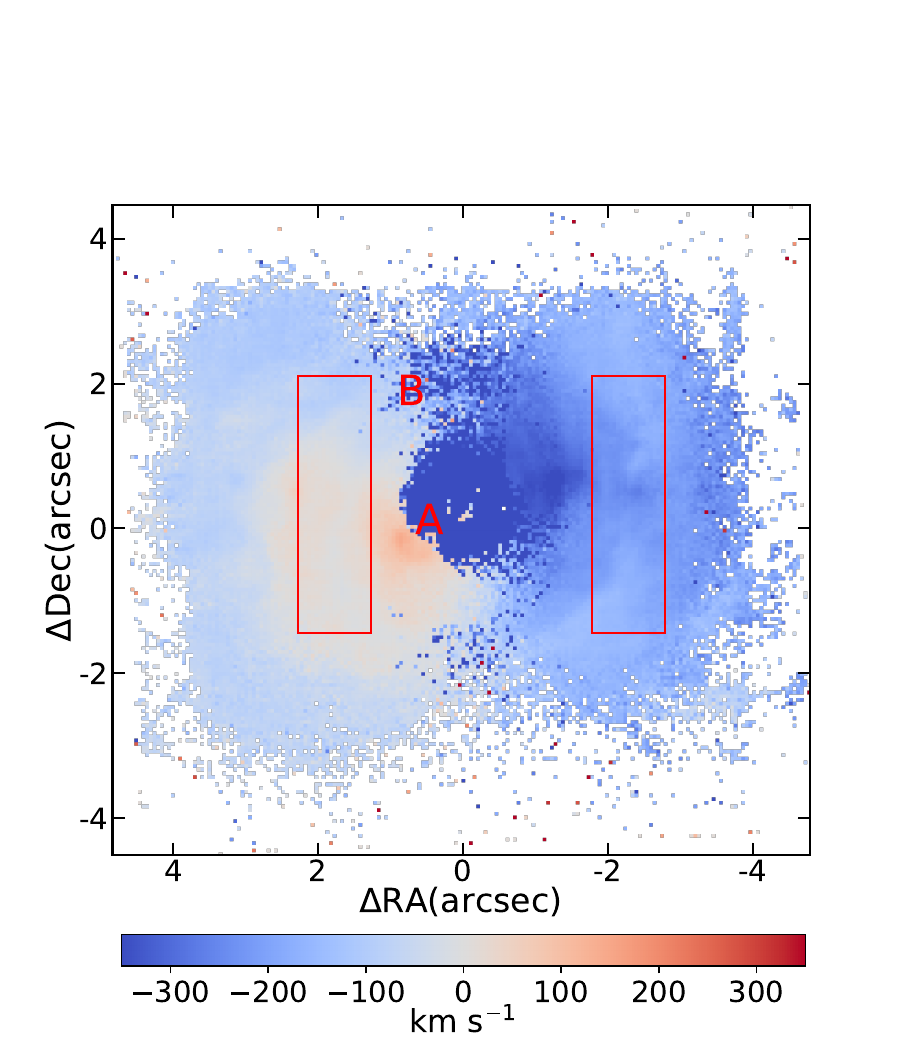}  
        \put(-195, 205){\textbf{(a)}}  
    \end{minipage}
    \hspace{0\textwidth}  
    \begin{minipage}{0.45\textwidth}
        \centering
        \includegraphics[width=\textwidth,trim={0 0 0 90}, clip]{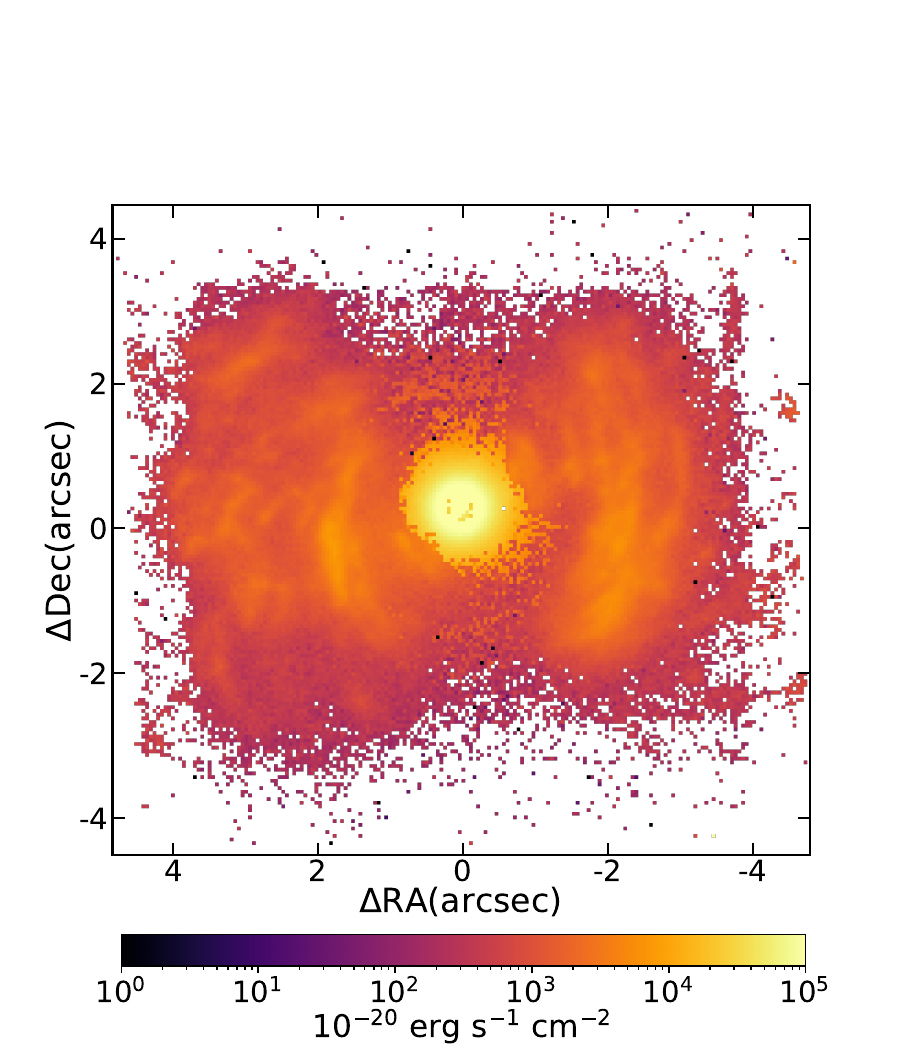}  
        \put(-195, 205){\textbf{(b)}}  
    \end{minipage}

\caption{Velocity and flux maps of the H$\alpha$ emission line for MR 2251-178. (a) The line center velocity map of  H$\alpha$ narrow component. (b)The flux map of H$\alpha$ narrow component. Only spaxels where the peak of the H$\alpha$ line is detected with SNR $>$ 5 are plotted. The red boxes indicate regions used as references for the selection.}
\label{fig2}

\end{figure*}

\section{The ELFO algorithm}
\label{elfo}
In fitting the emission lines from IFS data, each spectrum is typically fitted individually \citep{2025ApJS..277...57Z}. In this case, the emission line fitting for each spectrum is performed independently, and the initial guesses for the fitting parameters are fixed. To simplify data analysis, we binned the spectral data by summing the spectra of every four neighboring spaxels. Figure \ref{fig2} shows an example of the emission line fitting results for MR 2251-178, including the velocity and flux maps of the H$\alpha$ emission line. It is observed that the central region (``A'' in Figure \ref{fig2}), as well as the northern and southern regions of the map (``B'' in Figure \ref{fig2}), contain some strange spaxels. The central region exhibits a significant ultra blueshift throughout, which may be due to the high brightness of the broad component, making it difficult to decompose the narrow component of the H$\alpha$ line (This is also evident from Figure \ref{fig8}, where the velocity of the narrow component tends to align with that of the broad component when the latter is too strong, resulting in a large-scale blueshift in the central narrow line velocity map). The results from the northern and southern regions show spatially incoherent features, likely resulting from low SNR.

The goal of ELFO is to improve the decomposition of emission lines. During the fitting process, the entire IFS data is fitted in a specific sequence, with initial guesses for each spectrum's emission line fitting derived from neighboring spectra. We can modify the fitting order to produce multiple results for each spectrum, from which the one exhibiting spatial smoothness is selected using our method. Further details are provided below.
\subsection{Fitting process}
\begin{figure}[h]

\centering
\includegraphics[width=\linewidth]{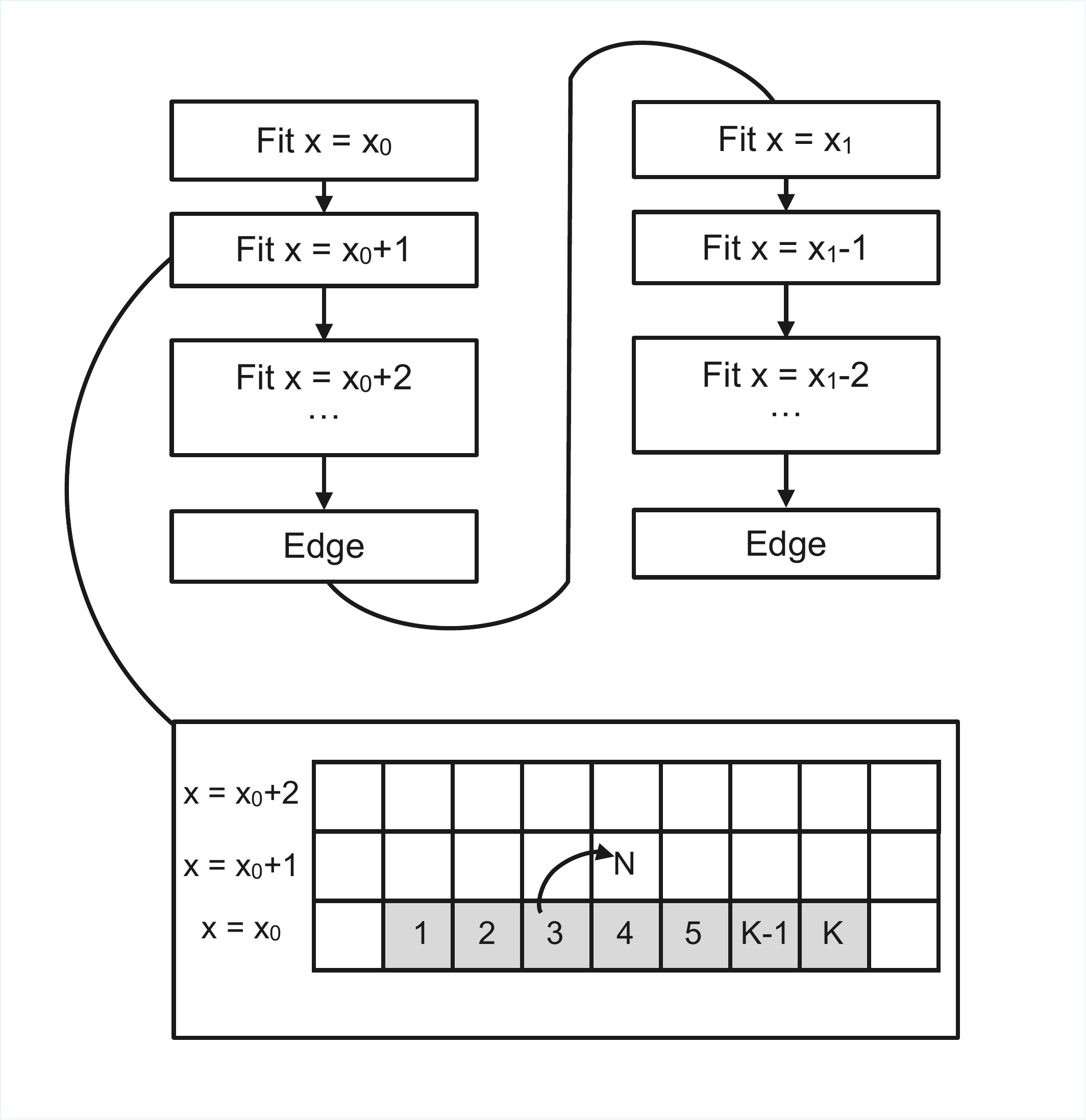}

\caption{Flowchart of the fitting process when $x_1 > x_0$. The lower inset illustrates the fitting process for row $x_0+1$. For each spectrum (spaxel N) in row $x_0+1$, the initial guess for the fitting is determined by the result with the minimum $\chi^2_\mathrm{red}$ among the $k$ neighboring spectra(spaxel 1, 2, 3\dots k) from row $x_0$.}
\label{fig3}

\end{figure}

In our fitting process, several parameters are defined by the user, the starting rows $x_0$, $x_1$, or the starting columns $y_0$, $y_1$, as well as the range $k$ of neighboring spectra. The flow of the fitting scheme is shown in Figure \ref{fig3}. We adopt the reduced chi-squared $\chi^2_\mathrm{red}$ as an estimate of the goodness of fit. Among the decomposition results of neighboring spectra, the one with the lowest $\chi^2_\mathrm{red}$ is considered the best-fit result. $\chi^2_\mathrm{red}$ is defined by:
\begin{equation*}
\chi^2_\mathrm{red} = \frac{1}{N - N_\mathrm{varys}} \sum_{i=1}^{N} \frac{(O_i - E_i)^2}{\sigma_i^2}
\end{equation*}

where: $ O_i $ is the observed value, $ E_i $ is the expected value (model), $ \sigma_i $ is the uncertainty of the observation, $ N $ is the number of data points, $N_\mathrm{varys}$ is the number of variables in fit, $ N - N_\mathrm{varys} $ is the degrees of freedom in fit.

\begin{figure}[h]

\centering
\includegraphics[width=1\linewidth]{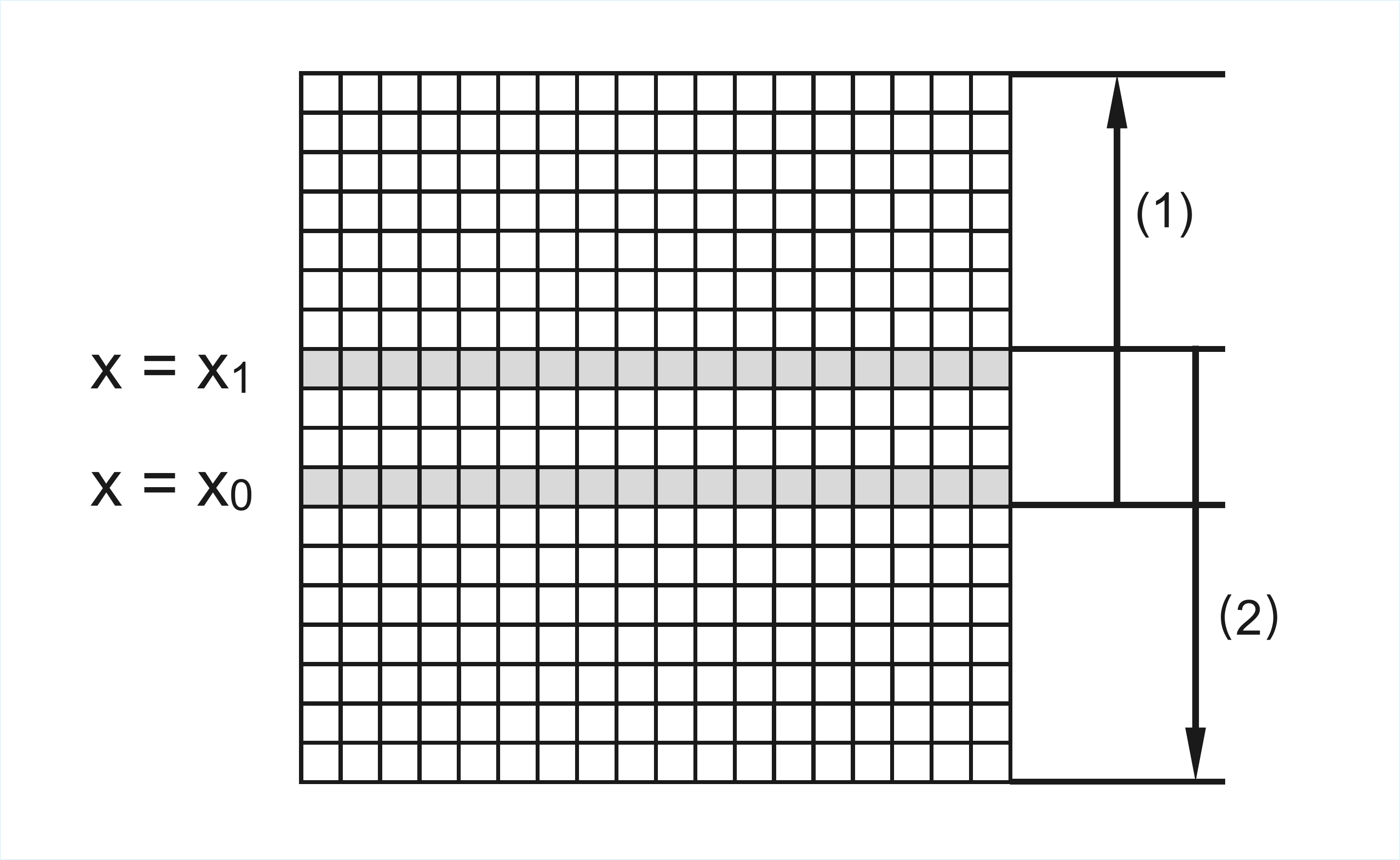}

\caption{Schematic diagram of the fitting sequence. The arrows indicate the fitting order: step (1) is executed first, followed by step (2).}
\label{fig4}

\end{figure}

Assuming the fit starts from row $x = x_0$
 , with $x_1>x_0$, the fitting of the emission lines for row $x = x_0$ is performed using the default initial guess values. Then, when fitting the specific spectrum in row $x_0+1$, the initial guess of the fitting is determined by the fit solution with the minimum $\chi^2_\mathrm{red}$ among the $k$ neighboring spectra from row $x_0$. This process continues for rows $x_0+2$, $x_0+3$, etc, until the fitting reaches the edge of the spectral data. Then the fitting process restarts from row $x = x_1$. Similarly, for each spectrum in row $x_1-1$, the initial guess for the fitting is determined by the fit solution with the minimum $\chi^2_\mathrm{red}$ among the $k$ neighboring spectra from row $x_1$. The fitting continues for rows $x_1-2$, $x_1-3$, and so on, until row 0. In other words, the process starts by fitting half of the spaxels from the middle and then fitting the other half from the opposite direction, as shown in Figure \ref{fig4}.
 
 An analogous approach is adopted when $x_1 < x_0$,  in which case the fitting proceeds from $x_0$ downward to row 0 for the first half and then from $x_1$ upward toward the opposite edge. The same logic applies when fitting along columns, with rows $x_0$ and $x_1$ replaced by columns $y_0$ and $y_1$. Appendix \ref{app1} explains why we adopt the above-described fitting procedure, fitting sequentially along rows and columns, instead of using the fit results from neighboring spectra of each individual spectrum as the initial guesses.

 After following the procedure described above, we will obtain a set of results. By setting multiple sets of parameters, each spectrum will have several results. The next step is to select the spatially smoothed result.

\subsection{Selection method}
\label{select}
\begin{figure}[h]

\centering
\includegraphics[width=0.8\linewidth]{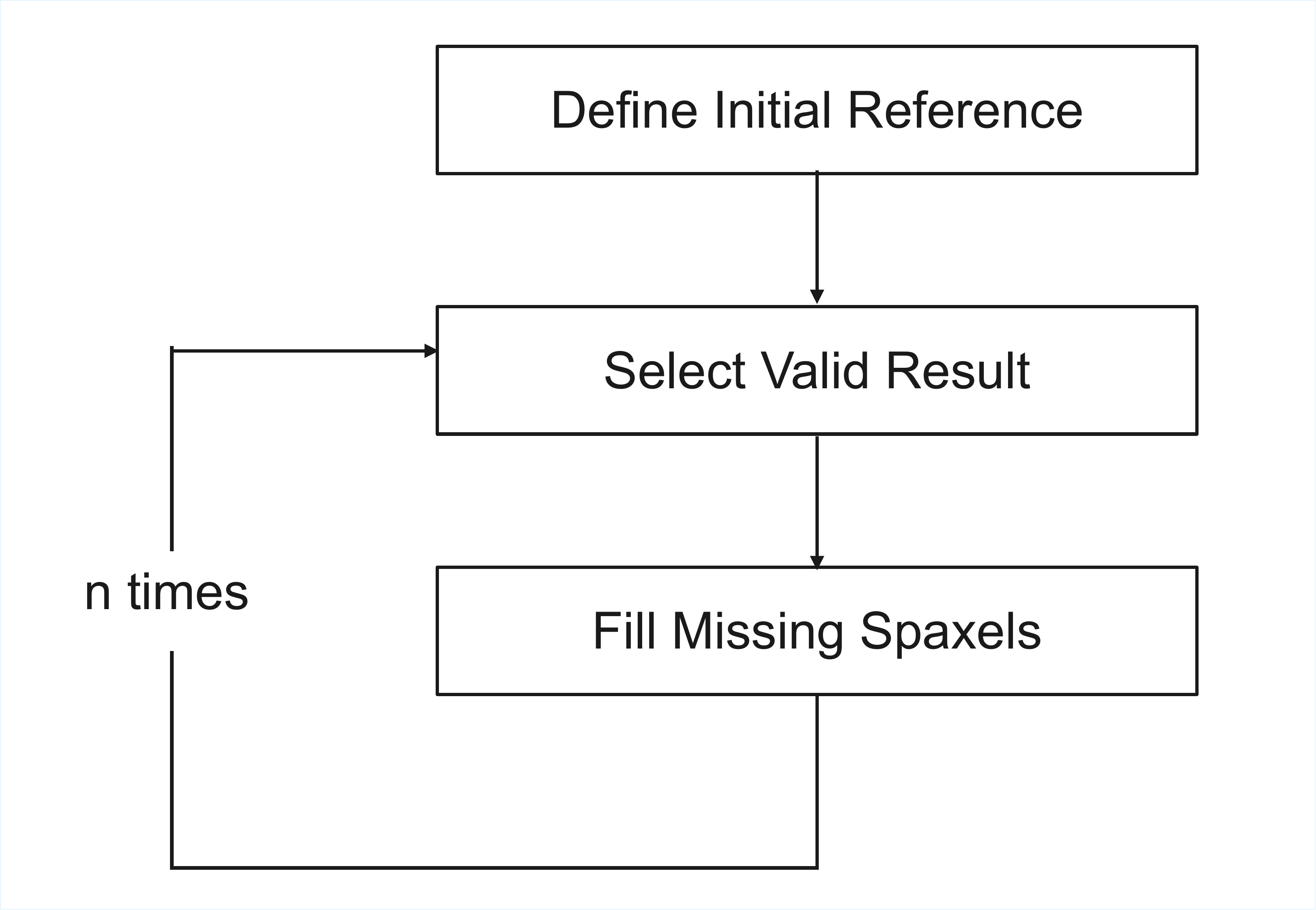}

\caption{Flowchart of the selection method. The selection process consists of four steps: (1) identifying the reference region; (2) selecting valid results; (3) filling in regions without reference values; and (4) repeating steps (2) and (3) until the set number of iterations is reached. For further details, refer to Section \ref{select}.}
\label{fig5}

\end{figure}

Once multiple fitting solutions have been obtained for each spectrum, the task becomes how to determine which one should be selected as the final result. Selecting the solution with the minimum $\chi^2_\mathrm{red}$ for each spectrum, as in the previous subsection, does not yield satisfactory results and often leads to unphysical discontinuities across the field. One possible reason is the large uncertainty of degrees of freedom for nonlinear models \citep{2010arXiv1012.3754A}, which undermines the reliability of reduced $\chi^2_\mathrm{red}$ as a comparative measure. We forgo seeking alternative statistical metrics, as they are likely to yield similar shortcomings. Instead, we adopt a different approach that exploits the spatial coherence among neighboring spectra. The basic assumption behind our method for selecting different results is that the velocity of the emission-line gas should vary smoothly. Figure \ref{fig5} outlines the selection process, which specifically consists of the following steps:

(1) In the initial velocity map, as shown in Figure \ref{fig2}, select the regions that are smooth and physically plausible (indicated by the red boxes in the figure). After removing spaxels with abrupt velocity changes, these selected spaxels are used as the reference for the initial selection. The median velocity dispersion of the H$\alpha$ narrow component from these spaxels is taken as $\sigma_0$.

(2) For each set of results, if the velocity of the H$\alpha$ narrow component differs from the reference within a range of $f_1\sigma_0$, and the velocity dispersion of the H$\alpha$ narrow component is smaller than $f_2\sigma_0$, the result is selected. Otherwise, it is excluded. Here, $f_1$ and $f_2$ are user-defined parameters. Recommended thresholds are $f_1 < 3$ and $f_2 < 5$ (e.g., $f_1 = 1.5$, $f_2 = 2$). Suppose that multiple results are selected for a single spaxel. In that case, the one with the smallest sum of absolute differences with the neighboring spaxels is chosen, i.e., we select the result that varies more smoothly on the plane.

(3) After obtaining the initial selection results, use the selected spaxels as the basis for further selection. For spaxels that were not selected, if more than three neighboring spaxels are selected, use the median of the neighboring velocity values as the selection criterion.

(4) Steps (2) and (3) are repeated until the preset number of iterations is reached.

This selection method may appear somewhat unconventional at first glance, but the results presented below clearly demonstrate its effectiveness. In fact, it is essentially equivalent to the intuitive, visual assessment of whether a given spectrum's fit is reasonable. The first step involves identifying visually smooth regions and excluding outliers or abrupt transitions. Steps (2) and (3) then proceed by expanding from these regions and selecting solutions for neighboring spaxels whose velocity differences vary smoothly without sudden changes. Normally, the velocity variation between two nearby spaxels should be comparable to the velocity dispersion at that location, making the threshold $f_1\sigma_0$ a physically reasonable criterion. The iterative application of steps (2) and (3) from the initially selected regions ensures that the final velocity field is spatially coherent.

\section{Results and discussion}
\label{sec4}
We employ ELFO to analyze the H$\alpha$ complex of MR 2251-178 and PG 1126-041. The line fitting parameters are shown in Table \ref{tab2}. For MR 2251-178, the parameter $k$ is set to 9, starting rows $x_0$, $x_1$ are : (83,93), (93,83), (76,66), (101,111). The first two sets of parameters are used for fitting starting from the central ten rows, while the latter two sets are applied from the top and bottom sides. For $y_0$, $y_1$, the settings are: (90,100), (100,90), (81,71), (109,119). Each set of parameters yields one set of fitting results. The final result is extracted from the eight sets of results through the selection method outlined in Section \ref{select}. 

Given the relatively low SNR of the PG 1126-041 data, we binned the spectral data by summing the spectra of every nine neighboring spaxels. The same procedure is applied to PG 1126-041, with the only difference being that the values for $x_0$, $x_1$, are set as: (54,64), (64,54), (53,43), (66,76), and the values for $y_0$, $y_1$ are set as: (55,65), (65,55), (53,43), (67,77)

\subsection{MR 2251-178}
\begin{figure*}[htp]

\centering
    \begin{minipage}{0.45\textwidth}  
        \centering
        \includegraphics[width=\textwidth,trim={0 0 0 90}, clip]{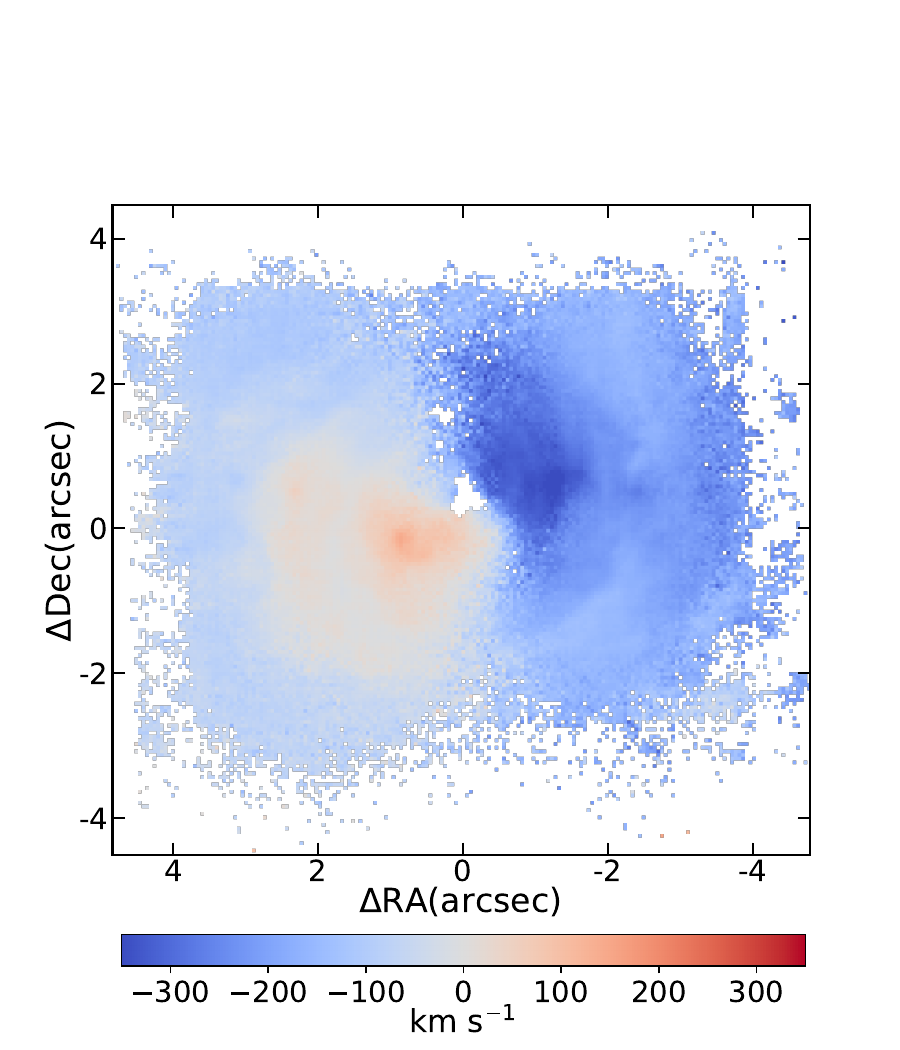}  
        \put(-195, 205){\textbf{(a)}}  
    \end{minipage}
    \hspace{0\textwidth}  
    \begin{minipage}{0.45\textwidth}
        \centering
        \includegraphics[width=\textwidth,trim={0 0 0 90}, clip]{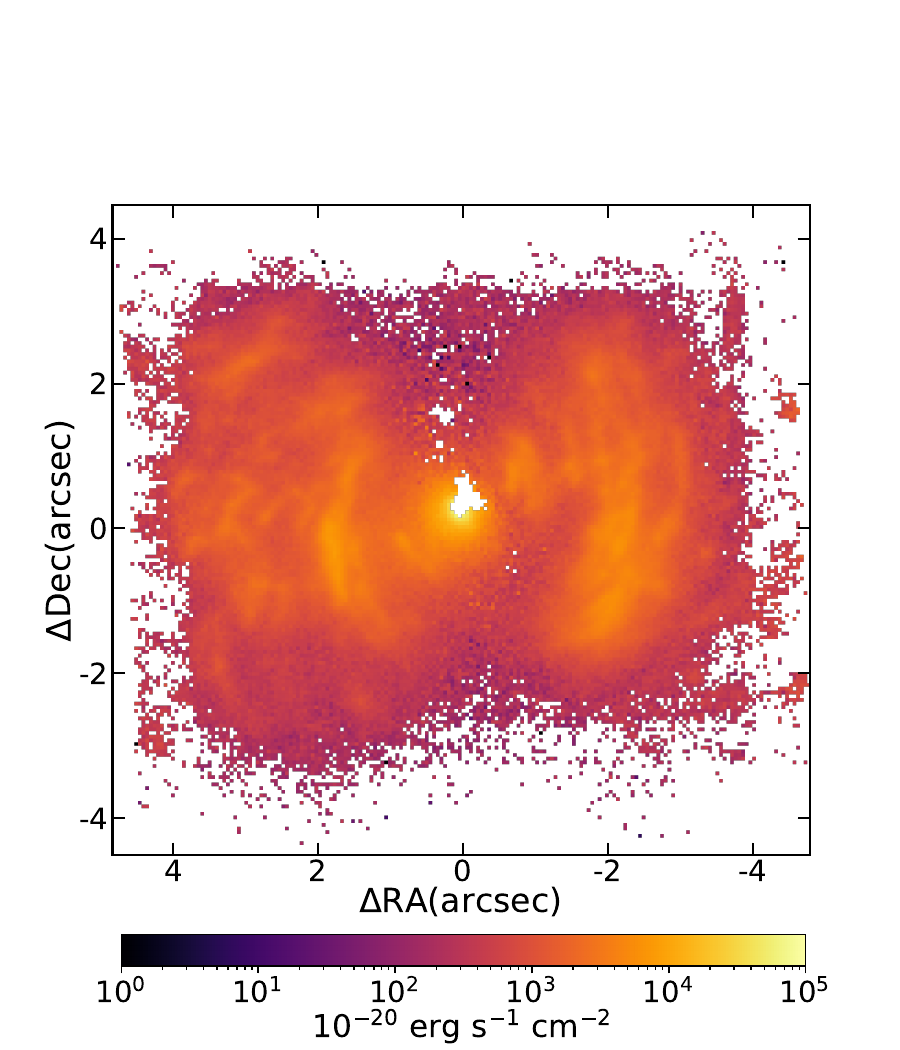}  
        \put(-195, 205){\textbf{(b)}}  
    \end{minipage}

\caption{Velocity and flux maps of the H$\alpha$ emission line for MR 2251-178, derived using the selection method. (a) The line center velocity map of  H$\alpha$ narrow component. (b)The flux map of H$\alpha$ narrow component. Only spaxels where the peak of the H$\alpha$ line is detected with SNR $>$ 5 are plotted.}
\label{fig6}

\end{figure*}

\begin{figure}[h]

\centering
\includegraphics[width=1\linewidth]{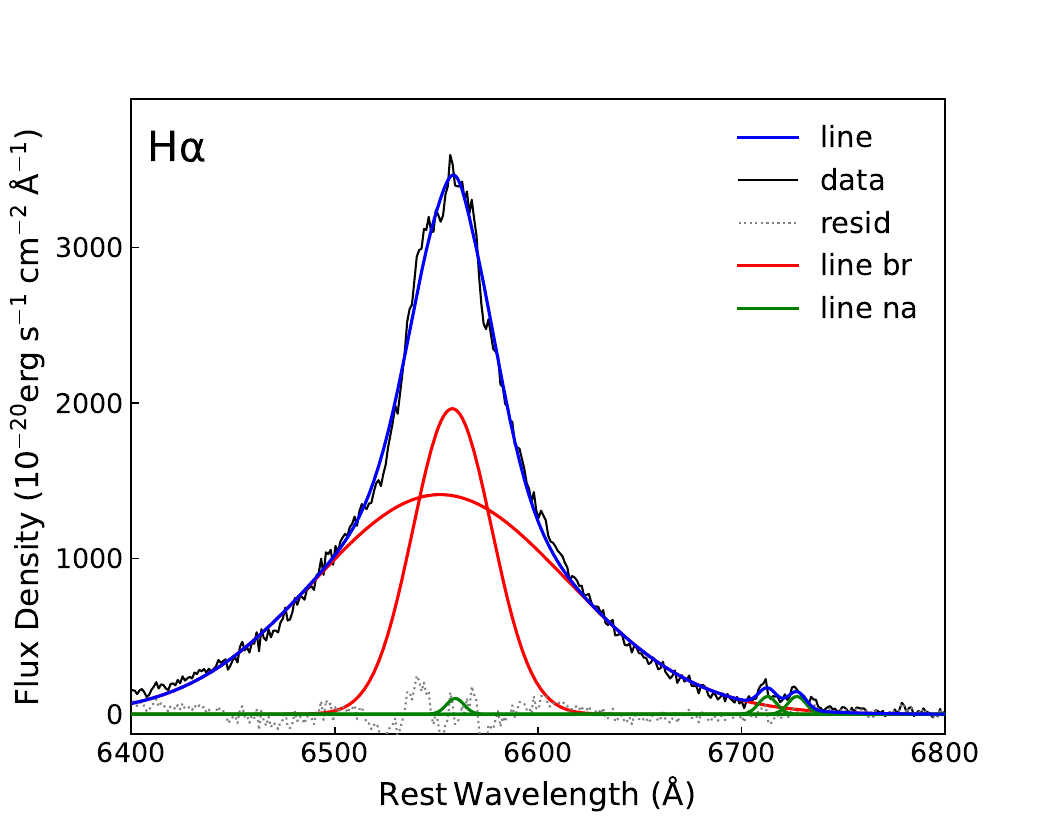}

\caption{Gaussian decomposition result for the H$\alpha$ emission line in the innermost region of MR 2251-178. The black solid line illustrates the observed data, while the blue solid line depicts the best-fitting model. The model consists of both broad (red solid line) and narrow (green solid line) emission lines. The gray dashed line represents the residuals of the emission line fitting.}
\label{fig7}

\end{figure}

Figure \ref{fig6} presents the final maps of MR 2251-178, including the velocity and flux maps of the H$\alpha$ emission line, selected using the method described in Section \ref{select}.  In the central region of the quasar, a small fraction of spaxels are not adopted after applying our selection method. This suggests that the fitting results in these areas exhibit abrupt deviations compared to their neighbors. The H$\alpha$ line profile near the central region, as shown in Figure \ref{fig7}, reveals that the H$\alpha$ emission is predominantly contributed by the broad component, with only a minor contribution from the narrow line component. As a result, the fitting model adopted in Table \ref{tab2}, one Gaussian for the narrow component and two Gaussians for the broad component, is not appropriate in this case.

\begin{figure}[h]

\centering
\includegraphics[width=\linewidth]{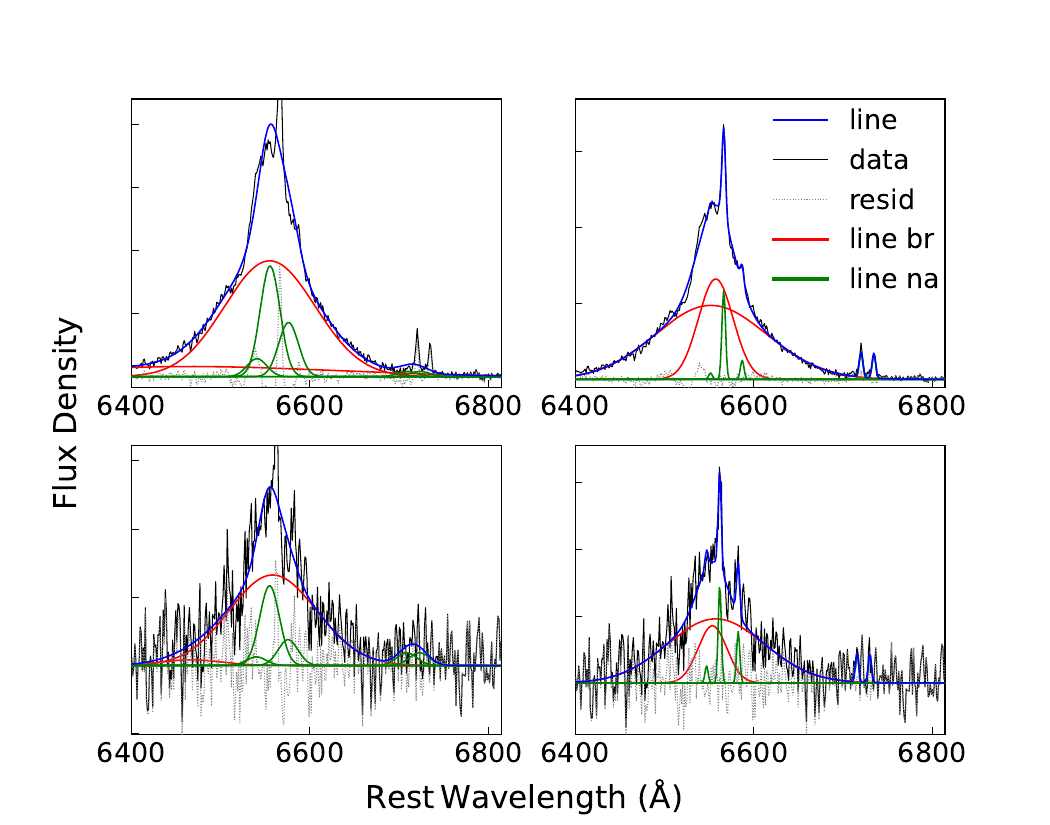}

\caption{Examples of Gaussian decomposition results for the H$\alpha$ emission line in MR 2251-178 before and after improvement. The left column shows the fitting results obtained with default Gaussian parameter guesses, and the right column displays the results selected by our algorithm. The first row corresponds to spectra from the central region (``A'' in Figure \ref{fig2}), and the second row shows spectra from the outer region (``B'' in Figure \ref{fig2}). The black solid line illustrates the observed data, while the blue solid line depicts the best-fitting model. The model consists of both broad (red solid line) and narrow (green solid line) emission lines. The gray dashed line represents the residuals of the emission line fitting.}
\label{fig8}

\end{figure}

Compared to Figure \ref{fig2}, the velocity map not only appears smoother in the northern and southern regions (``B'' in Figure \ref{fig2}), but also shows significant improvements in the central region (``A'' in Figure \ref{fig2}). Figure \ref{fig8} displays the initial and final emission-line decomposition results of the H$\alpha$ complex for these two cases (first row shows spectra from ``A'' in Figure \ref{fig2}, second row shows spectra from ``B'' in Figure \ref{fig2}). As can be seen, more reasonable Gaussian component decomposition results have been obtained in both cases.  It is important to note that the only modification made is setting the initial guesses informed by the results from neighboring spectra, followed by selecting the appropriate one. No smoothing or similar processing is applied to the velocity map shown in Figure \ref{fig6}.

\subsection{PG 1126-041}

\begin{figure*}[h]

\centering
    \begin{minipage}{0.45\textwidth}  
        \centering
        \includegraphics[width=\textwidth,trim={0 0 0 100}, clip]{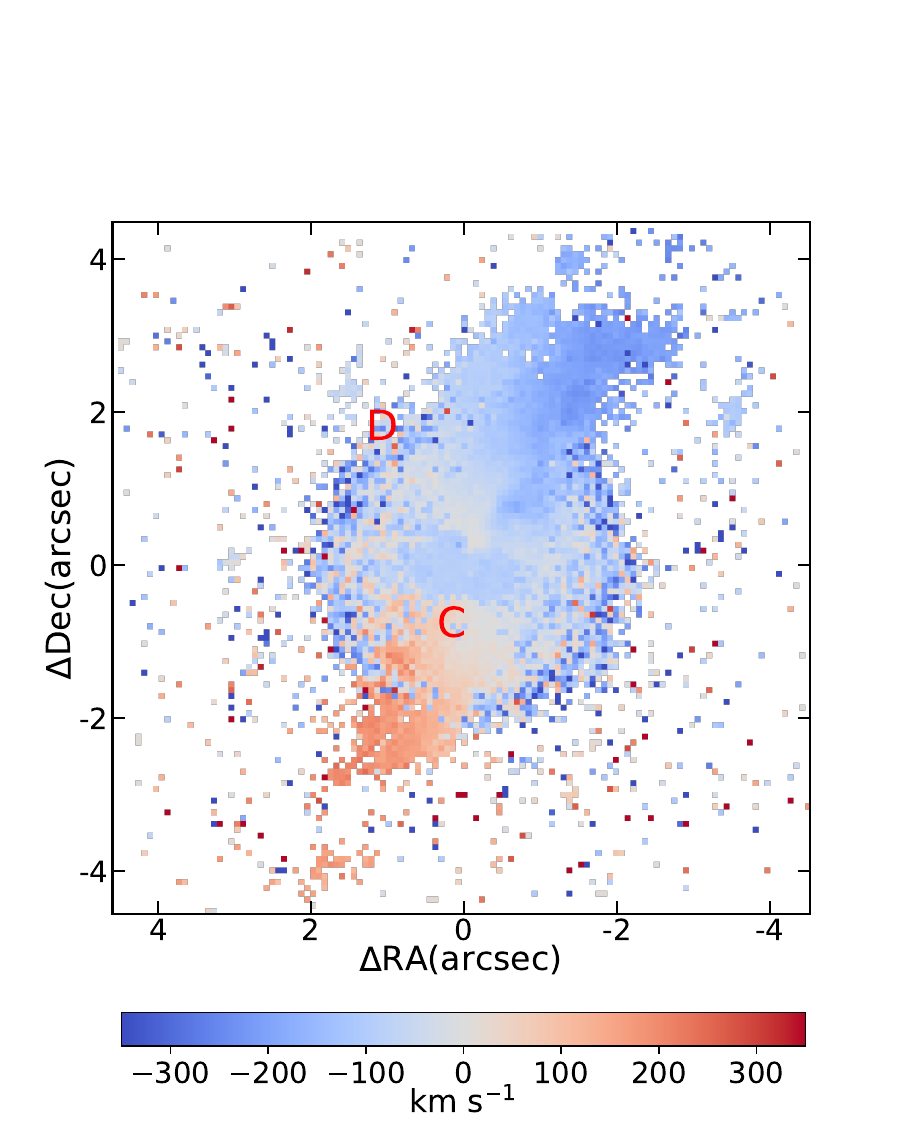}  
        \put(-195, 220){\textbf{(a)}}  
    \end{minipage}
    \hspace{0\textwidth}  
    \begin{minipage}{0.45\textwidth}
        \centering
        \includegraphics[width=\textwidth,trim={0 0 0 100}, clip]{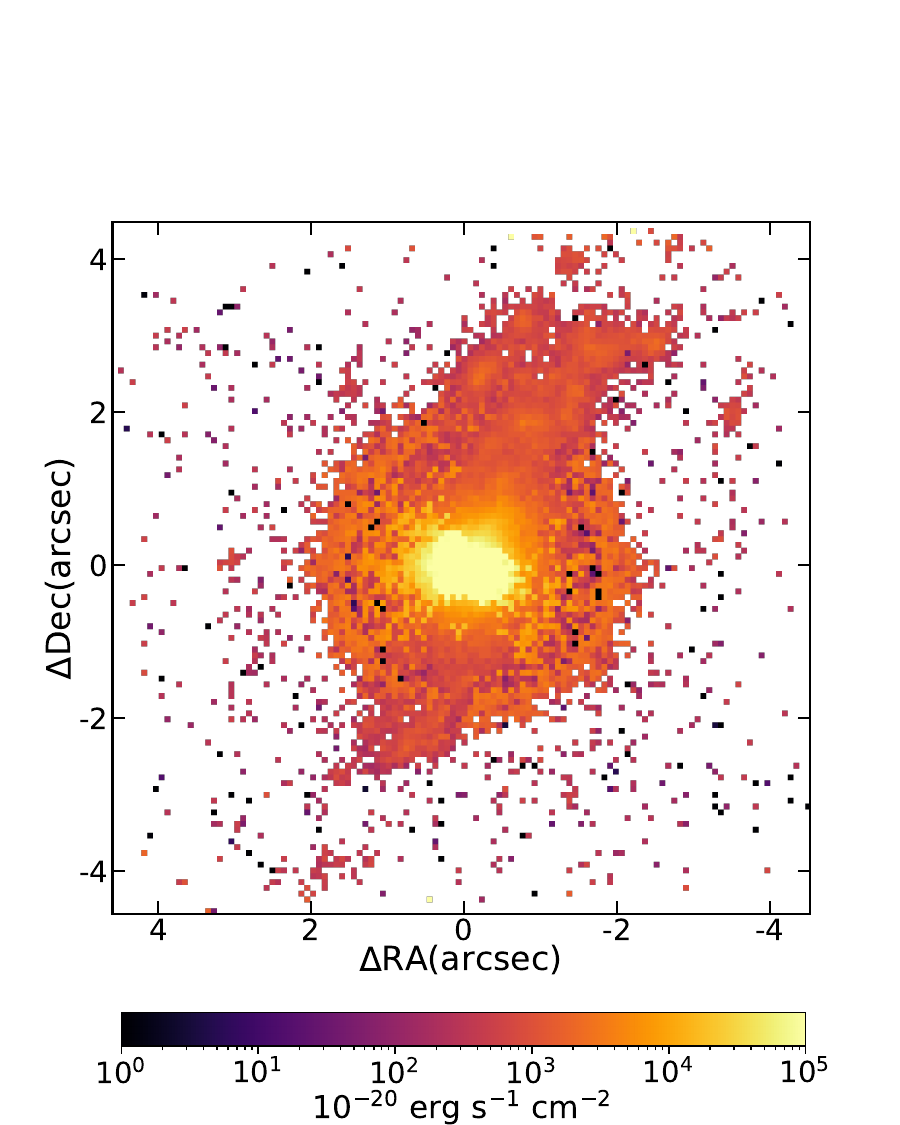}  
        \put(-195, 220){\textbf{(b)}}  
    \end{minipage}

\caption{Same as Figure \ref{fig2}, but for PG 1126-041}
\label{fig9}

\end{figure*}

\begin{figure*}[h]

\centering
    \begin{minipage}{0.45\textwidth}  
        \centering
        \includegraphics[width=\textwidth,trim={0 0 0 100}, clip]{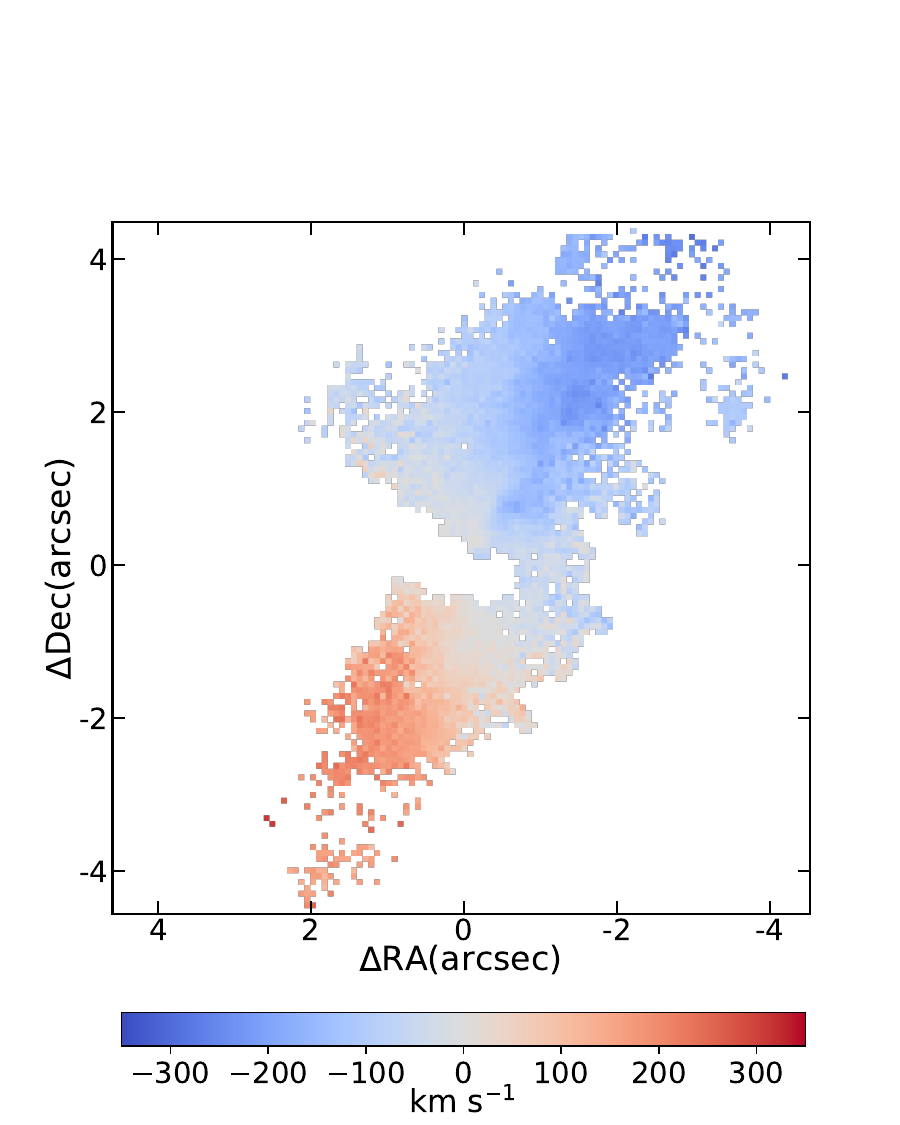}  
        \put(-195, 220){\textbf{(a)}}  
    \end{minipage}
    \hspace{0\textwidth}  
    \begin{minipage}{0.45\textwidth}
        \centering
        \includegraphics[width=\textwidth,trim={0 0 0 100}, clip]{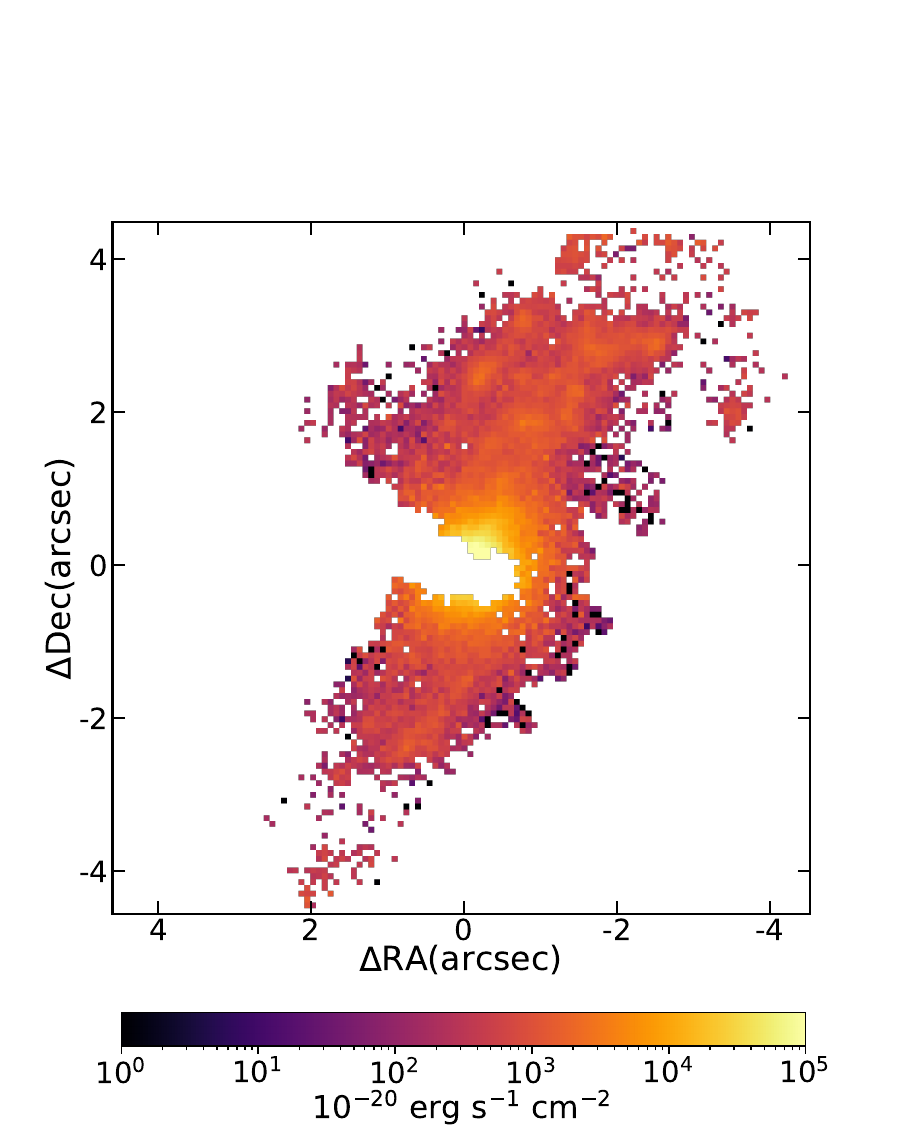}  
        \put(-195, 220){\textbf{(b)}}  
    \end{minipage}

\caption{Same as Figure \ref{fig6}, but for PG 1126-041}
\label{fig10}

\end{figure*}

The initial results for PG 1126-041, where each spectrum is fitted independently, are presented in Figure \ref{fig9}. The original velocity map shows numerous spaxels with significant errors, making the kinematic structure of the diffuse gas hard to interpret. By applying the procedure outlined in Section \ref{elfo}, we obtain the refined results shown in Figure \ref{fig10}.

\begin{figure}[h]

\centering
\includegraphics[width=\linewidth]{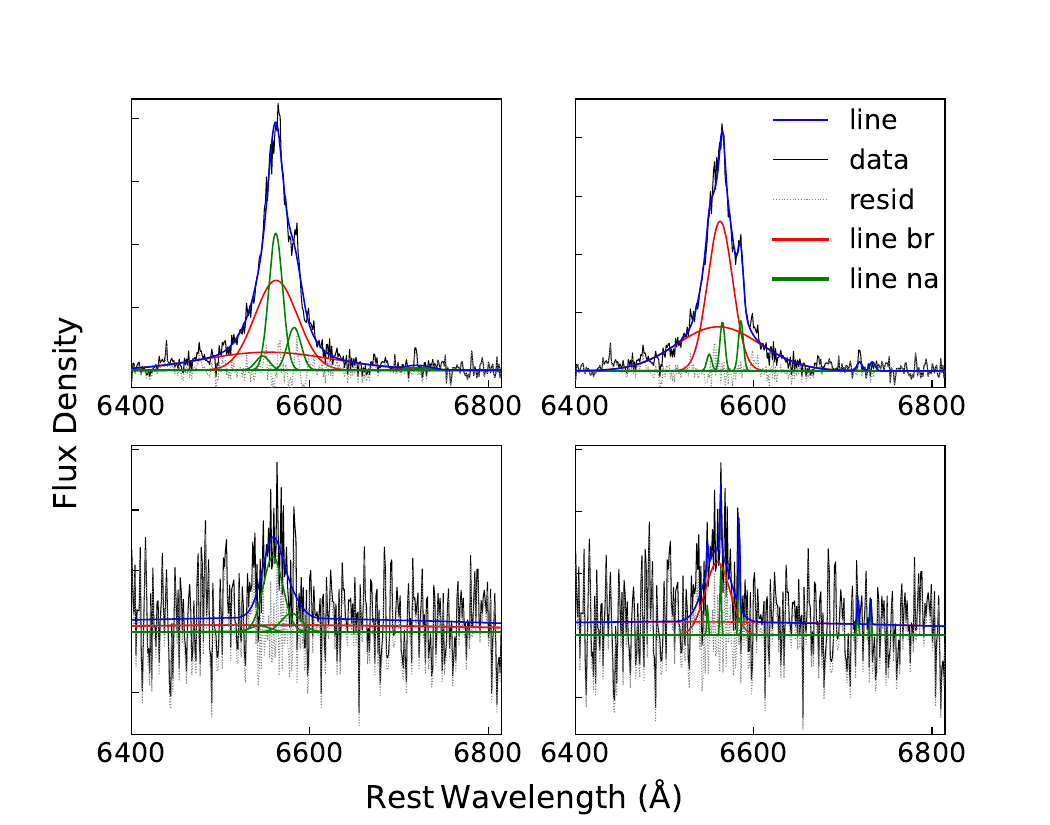}

\caption{Same as Figure \ref{fig8}, but for PG 1126-041. The first row corresponds to spectra from the central region (``C'' in Figure \ref{fig9}), and the second row shows spectra from the outer region (``D'' in Figure \ref{fig9}).}
\label{fig11}

\end{figure}

Analogously to MR 2251-178, a portion of spaxels near the quasar center is rejected by the selection criteria, which is attributed to the fact that the narrow line emission is negligible, making our model unable to reliably fit the observed line profile. Similarly to MR 2251-178, Figure \ref{fig11} displays the initial and final emission line decomposition results of the H$\alpha$ complex. The improvement for PG 1126-041 is more apparent as seen in clearer kinematics of the diffuse gas. Additionally, the structure that could not be clearly distinguished in Figure \ref{fig9} is now visible on the northeast side of Figure \ref{fig10}.


\subsection{Discussion}
In Marasco2020, to increase the SNR, a Voronoi adaptive binning \citep{2003MNRAS.342..345C} was employed. This approach helps reduce the complexity of subsequent spectral fitting, but also results in variations in the size and shape of the bins across areas of different SNR, and leads to significant information loss in the outskirts of the quasar. In contrast, by applying the ELFO algorithm combined with the selection method, we obtain smooth and reliable maps of the H$\alpha$ velocity and flux, with a uniform spaxel size across the entire field. This consistency enables a more accurate analysis of the kinematics and other properties of the ionized gas surrounding the quasar. 

The ionized gas surrounding MR 2251-178 has been extensively studied in previous works \citep{1983MNRAS.202..125B,1986A&A...169...49N,1999ApJ...524L..83S}. Early observations based on the H$\alpha$ and [\ion{O}{3}] emission lines revealed that the quasar is located within a very extended ionized nebula with a projected size of approximately 200 kpc. The origin of the ionized gas halo has been proposed to be a larger neutral envelope that is photoionized by the radiation field of the quasar \citep{1999ApJ...524L..83S}. The extended nebula rotates in the opposite direction to the gas near the quasar \citep{1986A&A...169...49N}. The MUSE data used in this study primarily cover the innermost region near the quasar.

Consistent with Figure 4 of Marasco2020, the final flux map (Figure \ref{fig6}) suggests that the nebula surrounding MR 2251-178 is composed of a series of east-west oriented thin shells of ionized gas. In addition, the velocity map shows a clear velocity gradient. However, in Marasco2020, due to the use of Voronoi adaptive binning, many spaxels near the edges of the field of view are grouped together, which smears out the kinematic structures in those regions. Consequently, the apparent velocity gradient is oriented along the southeast–northwest direction, which does not align with the east–west elongation of the flux. By contrast, in our final velocity and flux maps, the shell-like structures are evident in both. In the central region, the velocity gradient runs from southeast to northwest, matching the direction of the flux elongation, while the outer regions in both maps exhibit east–west oriented shell structures. The inconsistency between the directions of velocity gradient in the inner and outer regions suggests that if the shells are interpreted as a series of layers in spherical expansion from a central point \citep{2020A&A...644A..15M}, one would have to assume that the orientation of the quasar’s ionizing radiation cone has changed over time.

The ionized gas in PG 1126-041 exhibits relatively regular kinematics and morphology.  As shown in Figure \ref{fig10}, a clear northwest-southeast velocity gradient is observed, with a maximum velocity difference of less than 400 km s$^{-1}$. This gradient can be interpreted as arising from a rotating disk structure associated with the host galaxy. As the data for PG 1126-041 exhibit a lower SNR, the benefits from our optimization process are more evident.

In the analysis of emission lines for these two quasars, our algorithm demonstrates clear advantages. Our method can, on one hand, correct many anomalous decomposition results as seen in MR 2251-178, allowing the identification of previously unclear substructures; on the other hand, it can make the large-scale kinematic structure more evident in PG 1126-041. Our work illustrates that by reasonably setting the initial guess for fitting based on the correlation of neighboring spectra in IFS data, emission line fitting can be considerably optimized. Compared to commonly used adaptive binning techniques, this approach preserves more spatial information, facilitating subsequent analysis. However, the approach still requires iterative fitting over IFS spectra. For reference, on a personal Mac laptop equipped with the Apple M2 Pro chip, the fitting procedure for a single IFS dataset in this study requires approximately 5 hours when using five CPU cores in parallel. Repeating this task multiple times increases the total computation time accordingly. Therefore, the proposed algorithm is not recommended for preprocessing large volumes of IFS spectra and is best applied to detailed scientific analyses of individual targets or small samples. The development of ELFO was not tailored to any specific IFS instrument, and it can be conveniently employed for other IFS observations.

\section{Conclusion}
\label{sec5}
In this work, we present ELFO, a Python package for emission line fitting optimization in IFS data, based on \texttt{PyQSOFit} \citep{2018ascl.soft09008G}. We introduce the spectral fitting process and the principles of ELFO, as well as providing a detailed description of its fitting strategy and selection method. Additionally, we test ELFO on the MUSE data of two quasars.

In the analysis of emission lines for the two quasars, our method demonstrates clear advantages. By adjusting the initial fitting guesses, the method can correct anomalous fitting results,  identify previously unclear substructures, and make large-scale kinematic structures more evident. Compared to commonly used adaptive binning techniques, our approach preserves more spatial information. Our work shows that by reasonably setting the initial guesses for fitting based on the correlation of neighboring spectra and selecting according to the smooth variation of the velocity, the performance of emission line fitting can be significantly improved.

Although this paper has applied ELFO to only two MUSE data examples and focused on the H$\alpha$ emission line complex, the application of ELFO is not limited to these cases. The method has already been validated on CSST-IFS's simulated data \citep{2026RAA....26b4009F}\footnote{\url{https://csst-ifs-gehong.readthedocs.io}}, and with slight modifications, it can be applied to other IFS data and different emission lines. However, it should be emphasized that this approach requires multiple iterations of fitting on the IFS data, leading to increased computational time. Furthermore, this method, which relies on modifying initial guesses, cannot completely resolve the fitting difficulties caused by spectra with insufficient SNR.

In recent years, machine learning has been applied to nearly all fields of astronomy. The large data volume of IFS data makes it especially suitable for such techniques. During the development of our method, we found that in addition to modifying the initial parameters, constraining the allowed parameter ranges during the fitting process can significantly improve the results. However, setting appropriate bounds for each Gaussian component remains a challenge, and machine learning may offer a promising solution to this problem.

\section*{Acknowledgements}
We acknowledge the research grants from the Ministry of Science and Technology of China (National Key Program for Science and Technology Research and Development, No. 2023YFA1608100), the research grants from the China Manned Space Project (CMS-CSST-2025-A08), the National Natural Science Foundation of China (Nos. 12273036, 12222304, 12192220 and 12192221).

\appendix
\section{Assessment of Neighbor-Based Initialization Methods}
\label{app1}
When decomposing emission lines in spectra, performing a second round of fitting after the initial fit often yields improved results. As previously noted, parameters such as the gas velocity are expected to vary smoothly in extended targets. Moreover, the least-squares fitting of emission lines is sensitive to the choice of initial parameters. Therefore, it is natural to refit each spectrum using the results from neighboring spectra as an initial guess.

Several works have employed this strategy. For example, \citet{2016Ap&SS.361..280H} applied a spatial median smoothing to the original fit results across the entire data cube to generate new initial guesses for the refitting process - effectively taking the result from the neighboring spectrum, where the velocity is the median of the neighboring spectra, as the initial guess. Likewise, \citet{2022ApJ...935...72M} extracted new initial parameters from the best-fit models of neighboring spectra, retaining the solution with the lowest Bayesian Information Criterion (BIC). In another case, \citet{2019A&A...628A..78R} refitted flagged spectra by ranking their unflagged neighboring spectra according to their $\chi^2_\mathrm{red}$ values and using the results from the neighboring spectra with the lowest $\chi^2_\mathrm{red}$ as the initial guesses.

Although independently developed, our early attempts in this work also explored similar strategies. We found that adopting initial guesses based on the median or mean of the fitting results from neighboring spectra, or selecting the one with the lowest $\chi^2_\mathrm{red}$, could improve the fitting performance. Notably, even refitting the spectra using their own fitting results as the initial guesses yielded similar improvements. These strategies mitigate the issue of convergence to local minima caused by poor initial guesses. Furthermore, our tests with Mapping Nearby Galaxies at Apache Point Observatory(MaNGA) \citep{2015ApJ...798....7B} spectra showed that these approaches help to avoid erroneous fitting results arising from local spectral defects.

\begin{figure}[h]
\centering
\begin{minipage}{0.45\textwidth}
    \centering
    \includegraphics[width=\textwidth,trim={0 0 0 90}, clip]{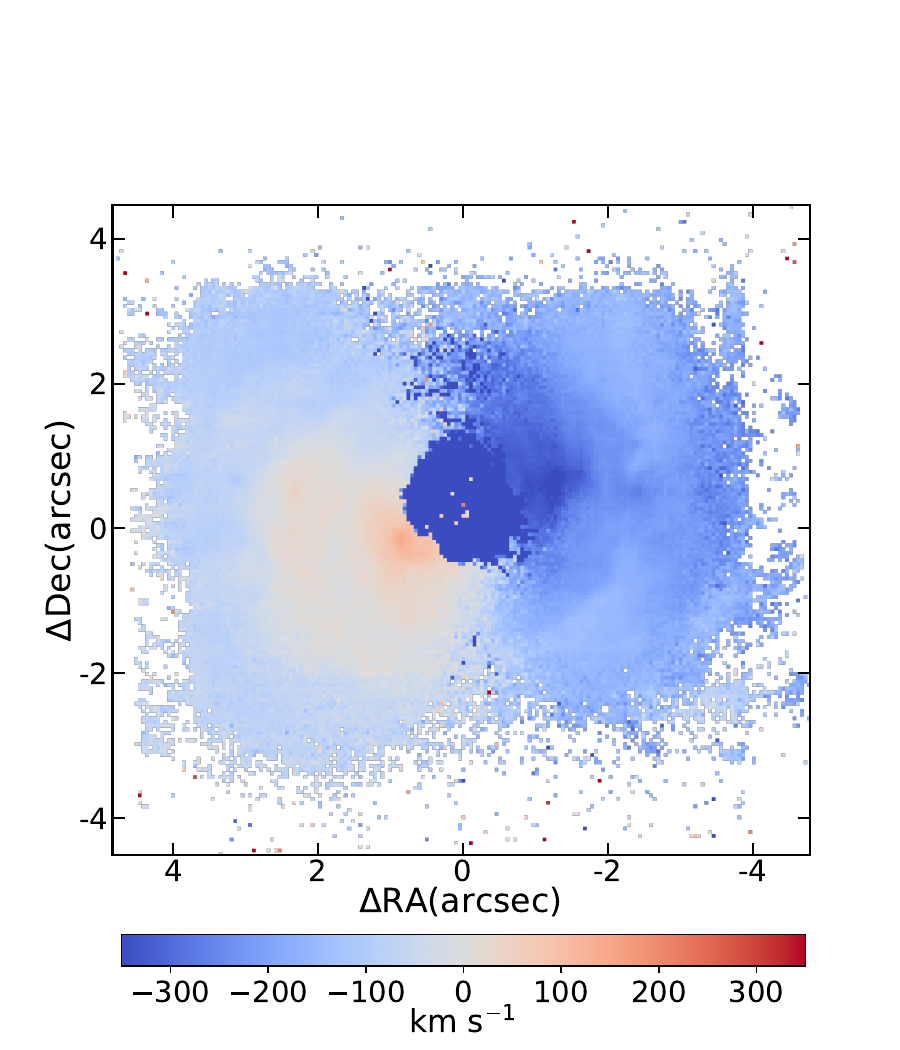}
    \put(-195, 205){\textbf{(a)}}  
\end{minipage}
\hspace{0\textwidth}
\begin{minipage}{0.45\textwidth}
    \centering
    \includegraphics[width=\textwidth,trim={0 0 0 90}, clip]{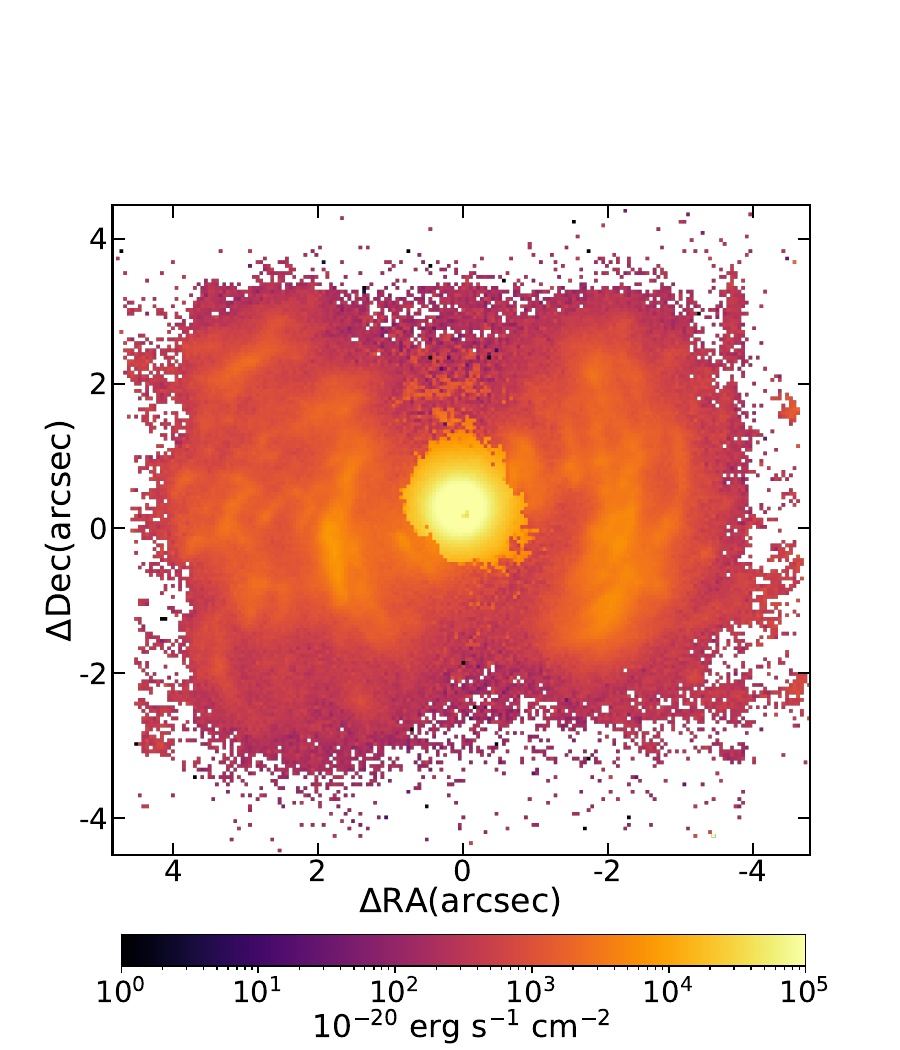}
    \put(-195, 205){\textbf{(b)}}
\end{minipage}

\vspace{0.05\textwidth}  

\begin{minipage}{0.45\textwidth}
    \centering
    \includegraphics[width=\textwidth,trim={0 0 0 90}, clip]{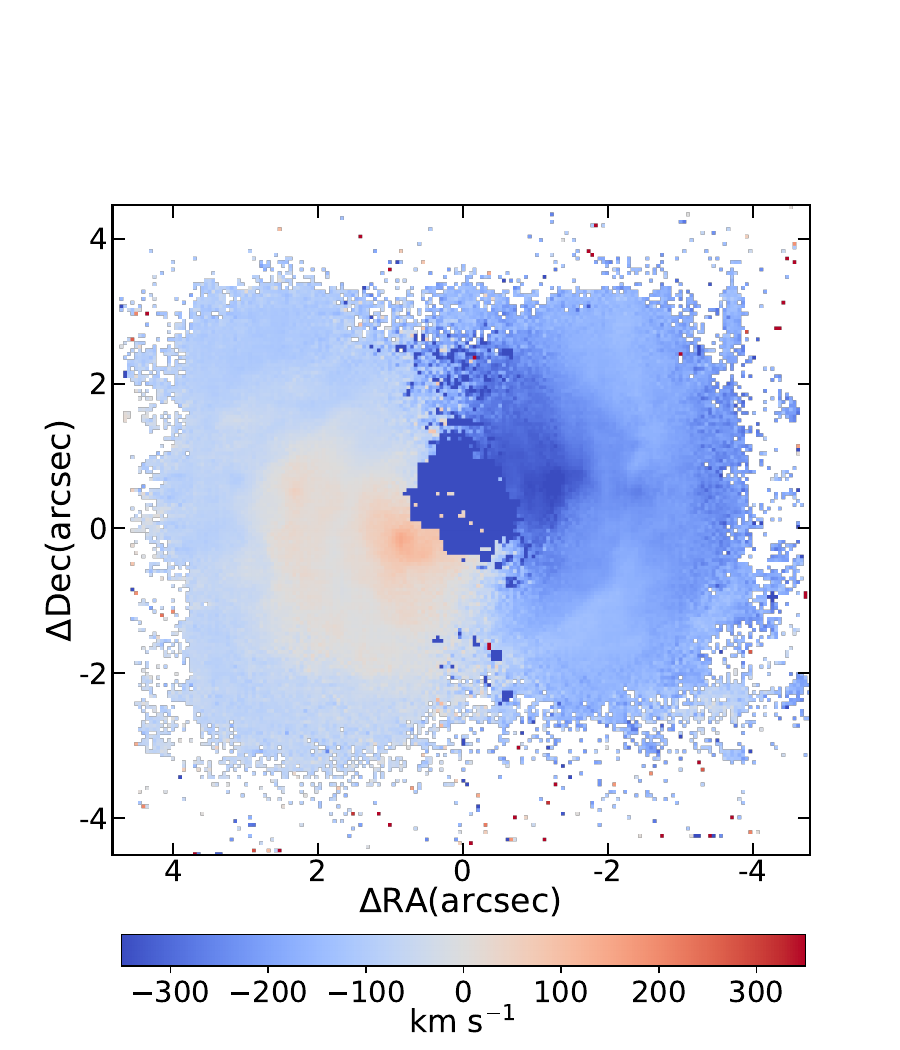}
    \put(-195, 205){\textbf{(c)}}
\end{minipage}
\hspace{0\textwidth}
\begin{minipage}{0.45\textwidth}
    \centering
    \includegraphics[width=\textwidth,trim={0 0 0 90}, clip]{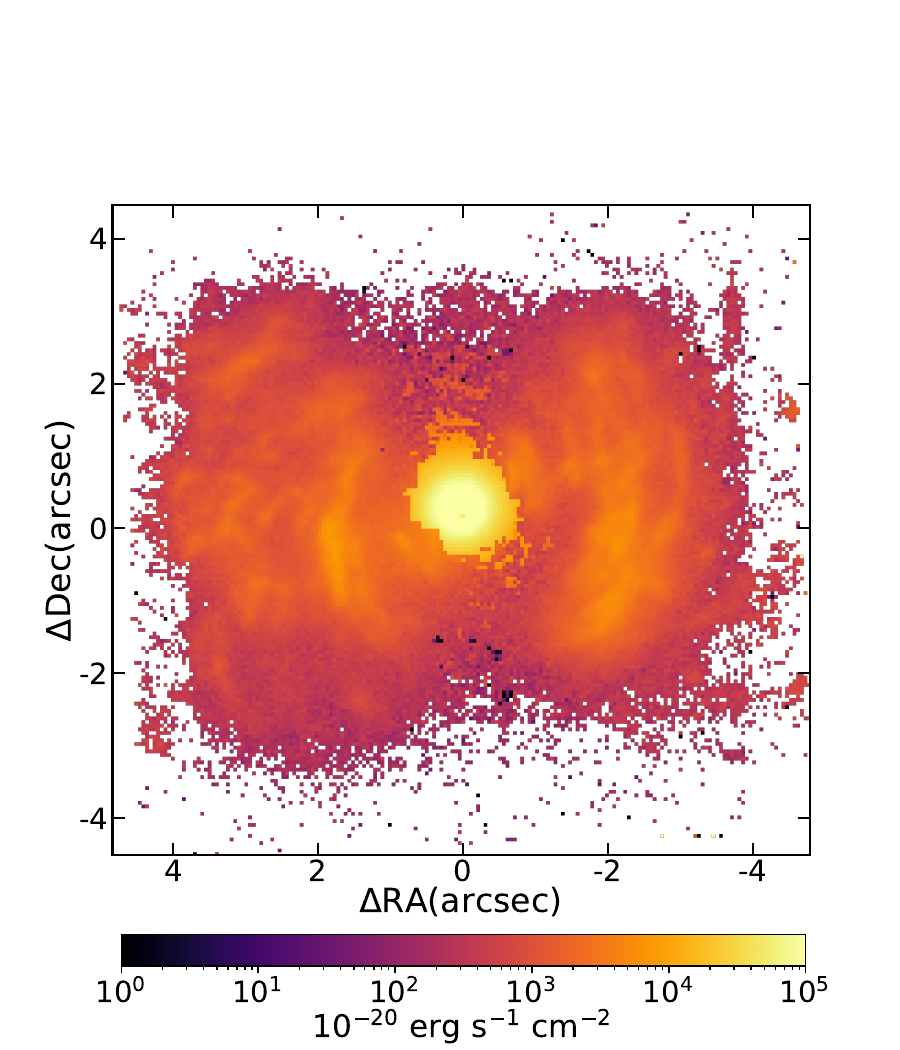}
    \put(-195, 205){\textbf{(d)}}
\end{minipage}

\caption{Refitted  Velocity and flux maps of the H$\alpha$ emission line for MR 2251-178, identical in content to Fig \ref{fig2}. The first row (a)(b) uses the result with the median velocity of all neighboring spectra as the initial guess for the emission line fit. The second row (c)(d) uses the result of the neighboring spectrum with minimum $\chi^2_\mathrm{red}$ as the initial guess.}
\label{fig12}

\end{figure}

However, when testing this strategy on MUSE data, we found that directly using the fitting results from neighboring spectra as initial guesses often led to only limited improvement in the fitting performance. Figure \ref{fig12} shows the results of refitting the MR 2251-178 spectra using either the neighboring result with median velocity or the one with minimum $\chi^2_\mathrm{red}$ as the initial guess for the emission line fitting. Compared to Figure \ref{fig2}, noticeable improvements can be seen in the northern and southern regions, consistent with what has been reported in the literature, but in the central region, if the initial fit exhibits a clear anomaly, neighboring spectra are also likely to be poorly fitted. As a result, this approach fails to significantly improve the fit solutions in these spectra. This motivates us to adopt the fitting strategy described in Section \ref{elfo}, in which spectra are recursively fitted along rows and columns. The fitting process can begin from any location within the field of view, which helps mitigate the aforementioned issues to some extent.

\bibliography{ref}{}
\bibliographystyle{aasjournalv7}

\end{document}